\documentclass[ajp,preprint]{revtex4-2}

\usepackage{amsmath}
\usepackage{amsfonts}
\usepackage{amssymb}
\usepackage{graphicx}
\usepackage{enumitem}
\usepackage{dcolumn}
\usepackage{bm}
\usepackage{braket}
\usepackage{ragged2e}
\usepackage{array}
\usepackage{tabularx}
\usepackage{hyperref}

\newcolumntype{Y}{>{\raggedright\arraybackslash}X}
\newcolumntype{P}[1]{>{\raggedright\arraybackslash}p{#1}}

\begin{document}
	
	\title{A Research-Informed Module on Quantum Superposition for Rapid Classroom Adoption}
	
	\author{Boris Kiefer}
\affiliation{Department of Physics, New Mexico State University, Las Cruces, NM 88003, USA}
	
	\begin{abstract}
		We present an adoption-ready instructional module for introducing quantum superposition in a two-state system. The package combines a five-activity classroom sequence with grading-ready assessment materials organized around six conceptual barriers documented in the physics education research literature: interpreting superposition as physical splitting, confusing coherent superposition with classical mixture, making basis-change errors, misreading finite-sample fluctuations as changes in the underlying state, using inconsistent notation, and, in an optional extension, reasoning about ordered operations. The main claim is that the bottleneck for introductory quantum instruction is rarely the absence of a usable simulator, but rather the absence of a coherent activity sequence, barrier-targeted prompts, and aligned assessment tools that an instructor can deploy without additional development work. We make the instructional rationale explicit through backward mapping from documented barriers to activity prompts and rubric-based evidence. The resulting module is designed for a single 50-minute class meeting and can be implemented with the included notebook or adapted to comparable two-state quantum simulators.
	\end{abstract}
	\maketitle
	
	\section{Introduction}
	
	Superposition is often the point at which students first confront the difference between classical state descriptions and amplitude-based quantum reasoning. Even in two-state systems, students commonly struggle to coordinate basis choice, state representation, and measurement statements \cite{Krijtenburg-Lewerissa2017-aj,Singh2015-qu,Marshman2015-um,Bouchee2022-nf}. Reported difficulties include treating superposition as physical splitting, confusing coherent superpositions with classical mixtures, making basis-dependent prediction errors, and misreading finite measurement samples as changes in the underlying state \cite{Marshman2017-hc,Zhu2012-mn,Passante2015-ll,Marshman2017-wg,Hu2023-fn}.
	
	A range of quantum learning tools already exists. PhET and QuVis lower access barriers through browser-based interactives and broad topic coverage \cite{Perkins2006-fr,kohnle2015-ho}, while research-oriented platforms such as QuTiP and Qiskit support flexible modeling in more advanced settings \cite{Johansson2012-uz,Javadi-Abhari2024-bp}. These tools address the technical barrier to simulation effectively. The instructional problem addressed here is different: an instructor preparing a single class meeting still must decide which conceptual barriers to address, write prompts that target those barriers, and construct aligned formative checks. This module does not claim simulator novelty as its primary contribution. It includes a reference simulator, but its main contribution is the activity sequence, concept checks, rubric, and implementation path that existing tools typically leave for the instructor to assemble.
	
	The module is intentionally narrow. Restricting scope to a two-state system with a fixed prepared input state allows the entire package to fit within a single class meeting. The payoff is a feasible implementation path, explicit alignment between documented difficulties and activity prompts, and assessment artifacts suitable for rapid grading. The finite-sampling activity deserves particular emphasis because it addresses a persistent misconception that is systematically underemphasized in short superposition modules: finite-run frequencies fluctuate even when the quantum state and theoretical probabilities remain unchanged \cite{Marshman2017-wg,Borish2024-lh}.
	
	To make the classroom use explicit, the instructor distributes a single worksheet containing the five activity prompts and the five concept-check items. Students alternate between short written predictions, simulator interaction, and brief explanations. After class, the instructor grades the collected worksheets with a rubric. The Supplementary Material provides the concept checks, pre/post prompts, answer key, deployment variants, and the reference notebook used in the paper. What is new here is not simulation capability, but the integration of barrier selection, activity wording, concept checks, rubric scoring, and a one-class implementation path into a single package ready for immediate use.
	
	\section{Design logic and barrier-to-design mapping}
	
	\subsection{Barrier selection}
	
	The module is organized around six recurrent learner barriers drawn from the PER literature. Barriers were included if they were documented in the literature and could be addressed within a two-state, fixed-preparation setting without formalism beyond introductory quantum mechanics. For the core in-class sequence (C1--C4, C6), we prioritized barriers supported by multiple independent studies. The ordered-operations extension (C5) is treated separately as an optional enrichment topic that is pedagogically accessible in this setting but not essential to the main 50-minute module. Barriers requiring density matrices, multi-qubit entanglement, or full Hamiltonian time evolution were excluded as outside the scope of a single class meeting, though they are natural extensions of this work.
	
	The six selected barriers are:
	\begin{itemize}[leftmargin=*]
		\item \textbf{C1:} superposition interpreted as physical splitting \cite{Marshman2017-hc,Passante2015-ll};
		\item \textbf{C2:} basis-change and measurement-context errors \cite{Zhu2012-mn,Hu2023-fn};
		\item \textbf{C3:} conflation of coherent superposition and classical mixture \cite{Passante2015-ll,Marshman2024-vi};
		\item \textbf{C4:} confusion between Born probabilities and finite-sample frequencies \cite{Marshman2017-wg,Borish2024-lh};
		\item \textbf{C5:} difficulty reasoning about ordered operations \cite{Emigh2015-nw};
		\item \textbf{C6:} notation-level ambiguity between amplitudes, basis states, and measurement claims \cite{Baily2010-yv,Baily2015-cu,Merzel2024-al}.
	\end{itemize}
	
	\subsection{Learning goals and backward mapping}
	
	These barriers motivate five learning goals. After the module, students should be able to: (LG1) distinguish coherent superposition from classical mixture; (LG2) predict basis-dependent measurement probabilities; (LG3) interpret finite-sample variability without changing the underlying state; (LG4) explain how phase conventions and basis choice affect quantum-state descriptions at an introductory level; and (LG5) communicate reasoning in consistent Dirac notation. Ordered-operation reasoning is treated as an optional extension aligned with LG4.
	
	Documented barriers determine the learning goals, which in turn determine the activity prompts and assessment evidence. Table~\ref{tab:per_to_design} makes that alignment explicit. An instructor adapting the module for a different topic can use the same structure: identify barriers from the literature, state the design response as a prompt or activity type, and specify which assessment item provides evidence for each learning goal.
	
	Barrier C6 is treated differently from the others. Rather than being assigned to a single activity, it is reinforced throughout the sequence via reflection prompts embedded in every activity and a dedicated rubric row. This choice reflects the finding that notation errors often persist even when conceptual understanding improves, and that targeted notation feedback at multiple points is more effective than a single isolated exercise \cite{Baily2010-yv,Merzel2024-al}.
	
	\begin{table*}[t]
		\caption{Mapping from documented student difficulties to design responses and assessment evidence. CC and PP labels refer to supplementary concept checks and pre/post prompts.}
		\label{tab:per_to_design}
		\footnotesize
		\begin{tabular}{@{}l@{\hspace{0.012\textwidth}}l@{\hspace{0.03\textwidth}}l@{\hspace{0.03\textwidth}}l@{\hspace{0.03\textwidth}}l@{}}
			\hline \\[-0.1cm]
			\parbox[t]{0.06\textwidth}{\textbf{Barrier}} &
			\parbox[t]{0.20\textwidth}{\textbf{PER finding(s)}} &
			\parbox[t]{0.20\textwidth}{\textbf{Design response}} &
			\parbox[t]{0.20\textwidth}{\textbf{Where implemented}} &
			\parbox[t]{0.25\textwidth}{\textbf{Learning goals and aligned evidence \\[0.2cm]}} \\
			\hline \\[-0.1cm]
			
			\parbox[t]{0.06\textwidth}{C1} &
			\parbox[t]{0.20\textwidth}{\justifying\noindent Superposition interpreted as splitting \cite{Marshman2017-hc,Passante2015-ll}.} &
			\parbox[t]{0.20\textwidth}{\justifying\noindent Amplitude-explicit labels and repeated Born sampling.} &
			\parbox[t]{0.20\textwidth}{\justifying\noindent Activities 1 and 3; simulator panel.} &
			\parbox[t]{0.25\textwidth}{\justifying\noindent \textbf{LG1, LG5:} \textbf{CC1}, \textbf{PP1}. Distinguish coherent superposition from classical mixture and explain single-shot versus ensemble language.} \\[2pt]
			
			\parbox[t]{0.06\textwidth}{C2} &
			\parbox[t]{0.20\textwidth}{\justifying\noindent Basis-tracking and measurement-context errors \cite{Zhu2012-mn,Hu2023-fn}.} &
			\parbox[t]{0.20\textwidth}{\justifying\noindent Explicit basis selector and side-by-side probability updates.} &
			\parbox[t]{0.20\textwidth}{\justifying\noindent Activities 2--4.} &
			\parbox[t]{0.25\textwidth}{\justifying\noindent \textbf{LG2, LG5:} \textbf{CC2}, \textbf{PP2}. Basis-specific projection prediction and notation-consistent justification.} \\[2pt]
			
			\parbox[t]{0.06\textwidth}{C3} &
			\parbox[t]{0.20\textwidth}{\justifying\noindent Mixture versus coherent superposition conflation \cite{Passante2015-ll,Marshman2024-vi}.} &
			\parbox[t]{0.20\textwidth}{\justifying\noindent Phase-sensitive comparison tasks in an incompatible basis.} &
			\parbox[t]{0.20\textwidth}{\justifying\noindent Activity 3.} &
			\parbox[t]{0.25\textwidth}{\justifying\noindent \textbf{LG1, LG4:} \textbf{CC1}, \textbf{PP1}. Identify coherence signatures in an incompatible basis; optional extension connects phase reasoning to ordered operations.} \\[2pt]
			
			\parbox[t]{0.06\textwidth}{C4} &
			\parbox[t]{0.20\textwidth}{\justifying\noindent Finite-sample misconceptions \cite{Marshman2017-wg,Borish2024-lh}.} &
			\parbox[t]{0.20\textwidth}{\justifying\noindent $N_{\text{trial}}$ control, histogram, and uncertainty language.} &
			\parbox[t]{0.20\textwidth}{\justifying\noindent Activity 5 and statistics panel.} &
			\parbox[t]{0.25\textwidth}{\justifying\noindent \textbf{LG3:} \textbf{CC3}, \textbf{PP3}. Separate sampling variability from changes in state preparation.} \\[2pt]
			
			\parbox[t]{0.06\textwidth}{C5} &
			\parbox[t]{0.20\textwidth}{\justifying\noindent Ordered-operation reasoning errors \cite{Emigh2015-nw}.} &
			\parbox[t]{0.20\textwidth}{\justifying\noindent Optional analytic extension on operator order.} &
			\parbox[t]{0.20\textwidth}{\justifying\noindent Supplementary challenge material.} &
			\parbox[t]{0.25\textwidth}{\justifying\noindent \textbf{LG4:} \textbf{CC4}, \textbf{PP4}. Introductory non-commutativity reasoning.} \\[2pt]
			
			\parbox[t]{0.06\textwidth}{C6} &
			\parbox[t]{0.20\textwidth}{\justifying\noindent Notation and interpretation disconnect \cite{Baily2010-yv,Merzel2024-al,Singh2015-qu}.} &
			\parbox[t]{0.20\textwidth}{\justifying\noindent Consistent notation and reflection prompts throughout.} &
			\parbox[t]{0.20\textwidth}{\justifying\noindent All activities and rubric rows.} &
			\parbox[t]{0.25\textwidth}{\justifying\noindent \textbf{LG5:} \textbf{CC5}, \textbf{PP5}. Coherent symbolic-to-verbal explanation quality. \\[-0.1cm]} \\
			\hline
		\end{tabular}
	\end{table*}
	
	\section{Quantum framework and simulator conventions}
	
	The activities assume a fixed prepared input state $\ket{+z}$ analyzed in a user-selected basis. The analysis basis is parameterized by the standard $U3$/ZYZ Euler decomposition \cite{Nielsen2011-xw}
	\begin{equation}
		U(\theta,\phi,\lambda) = R_z(\phi)\,R_y(\theta)\,R_z(\lambda),
	\end{equation}
	where $R_k(\varphi)=e^{-i\varphi\sigma_k/2}$. The displayed basis states are
	\begin{equation}
		\ket{0}=U^\dagger\ket{+z},\qquad \ket{1}=U^\dagger\ket{-z},
	\end{equation}
	and the Born-rule probabilities are
	\begin{equation}
		P_{\mathrm{theo}}(k)=|\braket{k|+z}|^2,\qquad k\in\{0,1\}.
	\end{equation}
	The convention is passive: $U$ rotates the analyzer frame rather than the state. Instructors using a tool that rotates the state vector, as in a Bloch-sphere display, can apply the same activity prompts with the substitution $U \to U^\dagger$ in the interface; all probability expressions and student-facing questions are unchanged. The full matrix form of $U(\theta,\phi,\lambda)$ is given in Supplementary Material S2 for reference.
	
	Several reference settings are useful across the activities: $U(0,0,0)$ gives the identity analyzer (aligned with the preparation), $U(\pi,0,\pi)$ gives the $\sigma_x$ analyzer, $U(\pi,\pi/2,\pi/2)$ gives $\sigma_y$, and $U(0,0,\pi)$ gives $\sigma_z$, all up to global phase \cite{Nielsen2011-xw}.
	
	A pedagogically useful feature of this parameterization is that different parameter triples $(\theta,\phi,\lambda)$ can produce basis states that differ only by overall phase. The printed ket components may change abruptly when angles wrap, even though all measurable probabilities remain unchanged. This provides a concrete, simulator-verifiable illustration of global-phase invariance that recurs across Activities~2 and~4.
	
\begin{figure*}[t]
		\centering
		\includegraphics[width=0.9\textwidth]{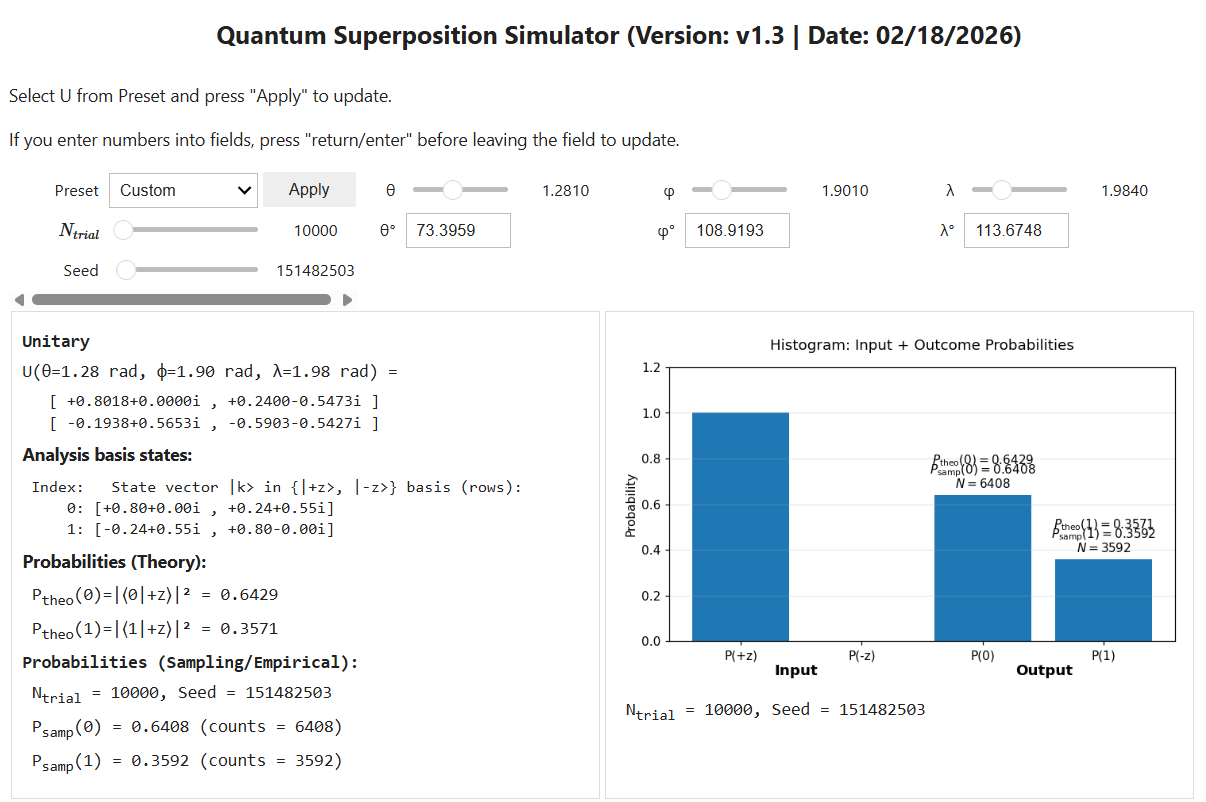}
		\caption{Simulator interface (included Jupyter notebook) showing analyzer-angle controls, trial count and seed inputs, theoretical and sampled probabilities, and an outcome histogram. Any two-state simulator with equivalent affordances can substitute; see text for portability notes.}
		\label{fig_sim}
	\end{figure*}

	The included Jupyter notebook is the reference implementation used in this paper because its notation and controls match the activity wording (Fig.~\ref{fig_sim}). The sequence can also be adapted to comparable two-state quantum simulators, including PhET or QuVis, provided that the interface displays basis states, reports Born-rule probabilities, and allows students to vary analyzer orientation and trial count. Minor wording changes may be needed when a tool uses different controls or a different visualization convention.
	
	\section{Activity sequence}
	
	The module is designed for a single class meeting, with optional challenge extensions assigned outside class. The five activities move from aligned-basis measurement through incompatible-basis reasoning to finite-sample statistics. Notation consistency is reinforced throughout via embedded reflection prompts rather than assigned to a single activity.
	
	Each activity uses a predict--observe--explain structure. Students first record a written prediction, then compare it with the simulator output, and finally write a brief explanation reconciling the two. Predictions are completed individually before simulator interaction, and the final written explanations are collected as the main assessment artifacts for that activity. This structure is consistent with evidence that recording written predictions before observation improves conceptual engagement and reduces uncritical acceptance of confirming outcomes \cite{crouch2001peer,kohnle2015-ho}. 
	
	\paragraph*{Activity 1: aligned-basis measurement (C1, C2).}
	With the analyzer set to the identity orientation ($\theta=0$), the basis is aligned with the preparation and the simulator reports $P_{\text{theo}}(0)=1$ and $P_{\text{theo}}(1)=0$. The prompt asks students to explain this outcome in terms of state--basis alignment rather than device behavior, and to predict what would change if the input state were $\ket{-z}$ instead. The purpose is to establish early that deterministic outcomes reflect a relationship between preparation and measurement basis, not a property intrinsic to either alone.
	
	\paragraph*{Activity 2: global phase and basis labeling (C1, C2).}
	Setting $\lambda=\pi$ with $\theta=0$ changes the sign of the displayed $\ket{1}$ component but leaves all probabilities unchanged. Students record both the displayed state vector and the probabilities, then explain in one sentence why the two descriptions are physically equivalent. This gives a compact, simulator-verifiable illustration of global-phase invariance and reinforces the distinction between displayed state vectors and observable outcomes --- a common source of notation confusion (C6).
	
	\paragraph*{Activity 3: incompatible basis and coherence (C2, C3).}
	This is the conceptual core of the module. With the analyzer set to the $x$ basis via $U(\pi/2,0,0)$, the simulator reports equal probabilities for the prepared state $\ket{+z}$ in both outcome slots. Students are then given two hypothetical source descriptions and asked to predict whether $x$-basis probabilities would differ between them. Source~A produces the coherent superposition
	\begin{equation}
		\ket{\psi_A} = \frac{\ket{+z}+\ket{-z}}{\sqrt{2}} = \ket{+x},
	\end{equation}
	while Source~B produces a 50/50 classical mixture of $\ket{+z}$ and $\ket{-z}$. Both sources yield identical $z$-basis statistics, so the $z$-basis simulator readout cannot distinguish them. Students project each description onto $\ket{+x}$ and $\ket{-x}$ and compare. The crucial observation is that Source~A is already the eigenstate $\ket{+x}$ and therefore gives a definite outcome in the $x$ basis, whereas Source~B gives equal probabilities. Class discussion links this result to the coherent-superposition versus mixture distinction and to the need to specify a basis when characterizing a state.
	
	\paragraph*{Activity 4: general basis exploration (C2).}
	Students vary $(\theta,\phi,\lambda)$ freely and record how the displayed basis states and probabilities change. The prompt directs attention to three observations: (i) probabilities always sum to one regardless of analyzer orientation; (ii) in this module's fixed-input convention, changing $\phi$ or $\lambda$ can alter the displayed phase labeling of the kets while leaving the reported probabilities unchanged when $\theta$ is fixed --- students are asked to write one sentence explaining why the probabilities are unaffected; and (iii) abrupt sign changes in displayed kets at angle wrap-around leave probabilities unchanged, revisiting the global-phase point from Activity~2. The ordered-operations extension is assigned only as a supplementary challenge so that the main sequence remains focused on basis and probability (Supplementary Material S3).
	
	\paragraph*{Activity 5: finite sampling and uncertainty (C4).}
	Students hold the analyzer fixed and vary only the trial count $N_{\text{trial}}$ and random seed, observing how the sampled frequency $\hat{p}$ fluctuates around the theoretical probability $P_{\text{theo}}$. The prompt asks them to estimate the standard error (Supplementary Material S4)
	\begin{equation}
		SE(\hat{p}) = \sqrt{\hat{p}(1-\hat{p})/N_{\text{trial}}}
	\end{equation}
	for two values of $N_{\text{trial}}$ and to explain explicitly why the spread in outcomes decreases without implying that the quantum state changed. This activity addresses an instructional gap common to short superposition modules: students often apply the Born rule correctly as a formula yet still interpret run-to-run variability as evidence of state collapse or state change \cite{Marshman2017-wg}. The emphasis on language, asking students to write a sentence that correctly attributes the variability to sampling, not to the state, connects naturally to C6.
	
	\begin{figure}[t]
		\centering
		\includegraphics[width=0.6\textwidth]{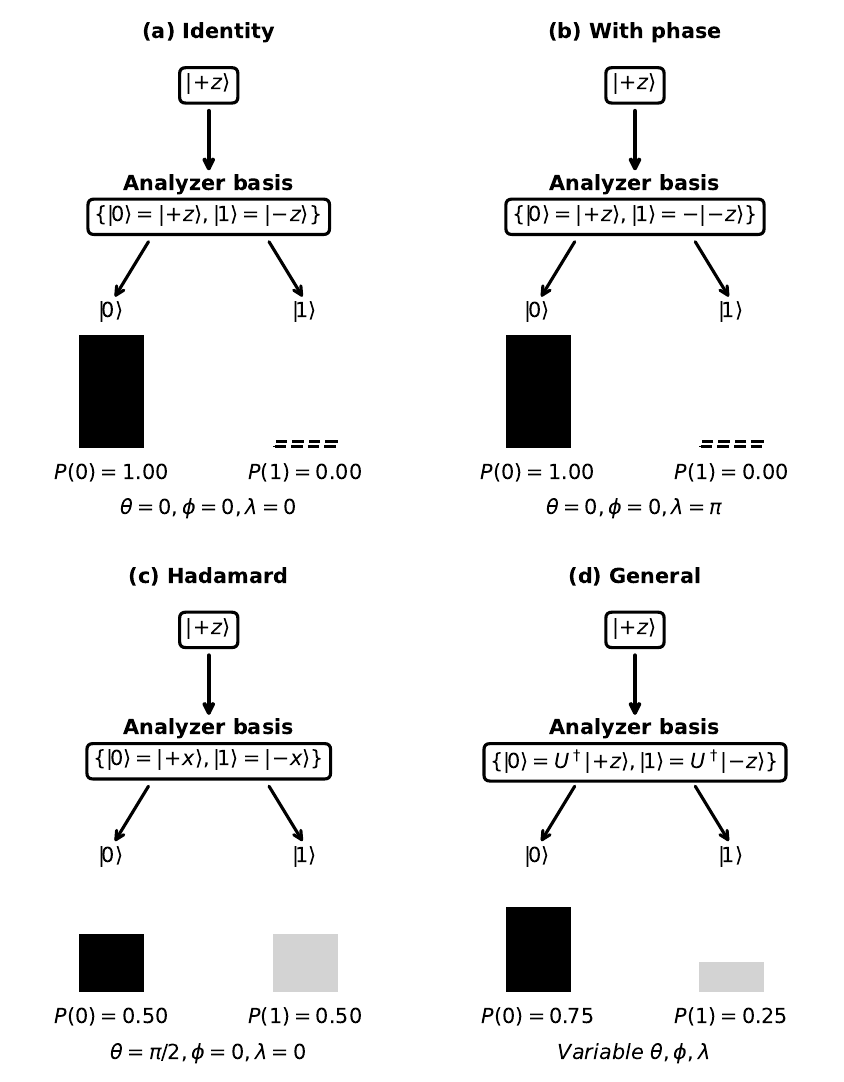}
		\caption{Measurement structure for Activities 1--4. The fixed prepared state $\ket{+z}$ is analyzed in basis states $\{\ket{0},\ket{1}\}=\{U^\dagger\ket{+z},U^\dagger\ket{-z}\}$, yielding probabilities $P_{\text{theo}}(k)=|\braket{k|+z}|^2$. Panel (a) shows aligned-basis measurement, (b) a phase-only display change, (c) incompatible $x$-basis measurement, and (d) a general analyzer orientation.}
		\label{fig21}
	\end{figure}
	
	\section{Implementation and assessment}
	
	\subsection{Timing and deployment modes}
	
	A full 50-minute implementation proceeds as follows: a 5-minute prediction-first warmup using the pre/post prompt PP1 (supplemental material S5); 10 minutes of instructor-led work on Activities~1 and~2; 12 minutes of paired work on Activity~3 including the pencil-and-paper prediction and class debrief; 10 minutes for Activity~4; 8 minutes for Activity~5; and a 5-minute exit prompt drawn from CC3 or CC5 (Supplementary Material S5). Optional challenge items may be assigned as homework.
	
	Two compressed modes are supported for time-constrained settings. In a \emph{partial} mode, Activities~1--3 are completed in class and Activity~5 plus one concept-check item are assigned as homework; this preserves the coherence/mixture distinction as the in-class centerpiece. In a \emph{flipped} mode, Activities~1 and~2 are completed before class using the pre-reading and simulator, so that class time focuses on Activity~3 and the statistics discussion in Activity~5.
	
	\subsection{Assessment design}
	
	Assessment is built at two levels: brief in-class concept checks (CC1--CC5) and matched pre/post prompts (PP1--PP5). All items are designed for paper collection and straightforward rubric-based grading without requiring a learning management system or automated scoring. Table~\ref{tab:rubric} summarizes the rubric, and the full item text and answer keys appear in the Supplementary Material S5.
	
	The rubric is designed for single-grader use in a typical course context. The binary LG5 criterion and the anchored partial-credit descriptions for LG1--LG3 are intended to minimize scorer discretion: each performance level is defined by the presence or absence of a specific identifiable element rather than by holistic impression. LG4 is optional and its rubric row applies only when the ordered-operations extension is assigned.
	
	\begin{table*}[t]
		\caption{Grading rubric for the module (10 points total).}
		\label{tab:rubric}
		\begin{tabular}{@{}ll@{}}
			\hline
			\textbf{Criterion} & \textbf{Performance levels} \\
			\hline
			
			\parbox[t]{1.35in}{LG1 (0--2 pts)} &
			\parbox[t]{5.4in}{\raggedright Superposition versus mixture: cites coherence/phase in an incompatible basis (2); partial basis reasoning (1); incorrect (0).} \\[4pt]
			
			\parbox[t]{1.35in}{LG2 (0--3 pts)} &
			\parbox[t]{5.4in}{\raggedright Probability prediction: correct projection setup and numerics (3); minor error or one probability correct (2); setup only (1); incorrect (0).} \\[4pt]
			
			\parbox[t]{1.35in}{LG3 (0--2 pts)} &
			\parbox[t]{5.4in}{\raggedright Sampling interpretation: correctly separates estimator variability from state change (2); partly correct (1); incorrect (0).} \\[4pt]
			
			\parbox[t]{1.35in}{LG4 (0--2 pts)} &
			\parbox[t]{5.4in}{\raggedright Optional operation-order reasoning: correct matrix products plus physical interpretation (2); partial argument (1); incorrect (0).} \\[4pt]
			
			\parbox[t]{1.35in}{LG5 (0--1 pt)} &
			\parbox[t]{5.4in}{\raggedright Notation consistency: consistent basis-aware notation throughout (1); inconsistent or incorrect (0).} \\
			
			\hline
		\end{tabular}
	\end{table*}
	
	Representative prompts illustrate the alignment between items and barriers. For LG1, students explain why a coherent superposition and a classical mixture can produce identical statistics in one basis but differ in another. For LG2, students compute $P(+x)$ and $P(-x)$ for a state expressed in the $z$ basis, showing the projection steps. For LG3, students interpret the effect of increasing $N_{\text{trial}}$ on the standard error without claiming that the state evolved or collapsed differently. These tasks keep the emphasis on basis-aware reasoning rather than algebra for its own sake. LG4 is included only as an optional extension and is not required for the core 50-minute implementation.
	
	\section{Scope and contribution}
	
	The module deliberately excludes density matrices, continuous-variable systems, multi-qubit entanglement, sequential projective measurement with explicit state update, and Hamiltonian time evolution. Each exclusion follows the same criterion used for barrier selection: the topic cannot be addressed within a two-state, fixed-preparation setting without introducing formalism that exceeds a single class meeting. These are natural next modules rather than weaknesses of this one.
	
	Within that scope, the paper contributes a classroom-ready example of barrier-targeted instructional design. The activity sequence, prompts, and rubric are tied to specific conceptual barriers documented in the PER literature rather than assembled as generic add-ons to a simulator. We treat the integrated design of activities, assessment, and implementation materials as a contribution independent of learning-gain measurement. The included assessment artifacts are ready for immediate classroom use and designed to support future validation work across institutional settings.
	
	\section{Conclusion}
	
	We have presented a classroom-ready superposition module for a two-state system built around documented student difficulties rather than around simulator features alone. Its main contribution is the coordinated packaging of activities, prompts, and assessment materials into a form that an instructor can adopt quickly for a single class meeting. The backward-mapping table makes the design logic explicit and portable, Activity~3 provides the conceptual centerpiece, and the finite-sampling activity addresses a difficulty that is often underrepresented in short introductory treatments of superposition.   
	
	\begin{acknowledgements}
		BK gratefully acknowledge support through the U.S. National Science Foundation under award numbers OST-2410813 and OST-2531569.
	\end{acknowledgements}
	
\section*{Data and code availability}
The simulator notebook, teaching materials, documentation, and example activities are available in the public GitHub repository at
\url{https://github.com/boriskiefer/sim_quantum_superposition}.
	
		\section*{author declarations}
	Conflict of Interest
	
	The author has no conflict to disclose.
	
	\bibliographystyle{unsrt}
	\bibliography{sim_superposition_ajp}

\clearpage

\section*{Supplementary Material}

\subsection*{S1. Deployment modes summary}

Three deployment modes are supported. In the \emph{full} 50-minute mode, instructors use all five activities in sequence as described in the main text. In the \emph{compressed} mode, Activities~1--3 are completed in class and Activity~5 plus one concept-check item are assigned as homework, preserving the coherence/mixture distinction as the in-class centerpiece. In the \emph{flipped} mode, Activities~1 and~2 are completed before class so that all class time can focus on Activity~3 and the statistics discussion in Activity~5. In all modes, the ordered-operations extension is assigned as an optional challenge outside class.

\subsection*{S2. $2\times 2$ unitary parameterization}

For reference, the full matrix form of the $U3$/ZYZ parameterization used in the simulator is
\[
U(\theta,\phi,\lambda)=
\begin{pmatrix}
	\cos\dfrac{\theta}{2} & -e^{i\lambda}\sin\dfrac{\theta}{2}\\[6pt]
	e^{i\phi}\sin\dfrac{\theta}{2} & e^{i(\phi+\lambda)}\cos\dfrac{\theta}{2}
\end{pmatrix},
\qquad
\theta\in[0,\pi],\ \phi,\lambda\in[0,2\pi).
\]
This is the standard Euler factorization of Nielsen and Chuang \cite{Nielsen2011-xw}.

\subsection*{S3. Optional extension: ordered operations}

For instructors who want a short extension on non-commutativity, comparing $H\sigma_x$ and $\sigma_x H$ provides an accessible entry point:
\[
H\sigma_x=\frac{1}{\sqrt{2}}\begin{pmatrix}1&1\\-1&1\end{pmatrix},
\qquad
\sigma_x H=\frac{1}{\sqrt{2}}\begin{pmatrix}1&-1\\1&1\end{pmatrix}.
\]
Applied to $\ket{+z}$, the two orderings produce orthogonal states. Since orthogonal states cannot differ by a global phase, the order is physically meaningful. This is sufficient to motivate the concept of non-commutativity without requiring the commutator formalism.

\subsection*{S4. Standard error for binary outcomes}

Let $X_i\in\{0,1\}$ denote the $i$th trial with $\mathbb{P}(X_i=1)=p$, and let $\hat{p}=N^{-1}\sum_i X_i$. Independence and $X_i^2=X_i$ give
\[
\mathrm{Var}(\hat{p})=\frac{p(1-p)}{N},
\qquad
SE(\hat{p})=\sqrt{\frac{p(1-p)}{N}}.
\]
This is the uncertainty estimate used in Activity~5 and CC3/PP3.

\subsection*{S5. Concept checks and pre/post prompts}

\noindent\textbf{CC1 (LG1).} Two sources produce identical 50/50 outcomes in the $z$ basis. Source~A is the coherent state $\ket{\psi}=(\ket{+z}+\ket{-z})/\sqrt{2}$. Source~B is a 50/50 classical mixture of $\ket{+z}$ and $\ket{-z}$. Are A and B physically equivalent? Justify your answer using an incompatible basis.

\medskip
\noindent\textbf{CC2 (LG2).} Given
\[
\ket{\psi}=\cos\!\left(\frac{\theta}{2}\right)\ket{+z}+e^{i\phi}\sin\!\left(\frac{\theta}{2}\right)\ket{-z},
\]
predict $P(+x)$ and $P(-x)$ and evaluate numerically for $(\theta,\phi)=(\pi/3,\pi)$.

\medskip
\noindent\textbf{CC3 (LG3).} One simulator run reports $\hat{p}=0.62$ from $N=100$ trials and another uses the same state and analyzer with $N=1000$. Compare the expected sampling uncertainty using $SE(\hat{p})=\sqrt{\hat{p}(1-\hat{p})/N}$. Does changing $N$ change the quantum state?

\medskip
\noindent\textbf{CC4 (LG4).} Compute $H\sigma_x$ and $\sigma_xH$, apply both to $\ket{+z}$, and explain why orthogonality of the resulting states rules out global-phase equivalence.

\medskip
\noindent\textbf{CC5 (LG5).} A student writes: ``Since $\ket{\psi}=a\ket{+x}+b\ket{-x}$, therefore $P(+z)=|a|^2$.'' Identify the error and give a correct basis-aware method.

\bigskip
\noindent\textbf{PP1 (LG1).} Explain in 2--4 sentences why a coherent superposition and a classical mixture can agree in one basis but disagree in another.

\medskip
\noindent\textbf{PP2 (LG2).} For an instructor-selected state, compute measurement probabilities in a specified basis and show the projection steps.

\medskip
\noindent\textbf{PP3 (LG3).} Two runs use the same analyzer setting but different $N_{\text{trial}}$. Explain why the observed frequencies differ and how the uncertainty scales with $N_{\text{trial}}$.

\medskip
\noindent\textbf{PP4 (LG4).} Under what conditions do two ordered products of unitaries produce physically distinct analyzers? Illustrate with one example.

\medskip
\noindent\textbf{PP5 (LG5).} Diagnose and correct a short notation inconsistency involving bras, kets, and probability extraction.

\subsection*{S6. Answer key}

\noindent\textbf{CC1 / PP1.} Full credit requires stating that a coherent superposition and a classical mixture can agree in one basis yet differ in an incompatible basis because the superposition retains phase coherence and the mixture does not.

\medskip
\noindent\textbf{CC2 / PP2.} Using $\ket{\pm x}=2^{-1/2}(\ket{+z}\pm\ket{-z})$,
\[
P(+x)=\tfrac{1}{2}(1+\sin\theta\cos\phi),
\qquad
P(-x)=\tfrac{1}{2}(1-\sin\theta\cos\phi).
\]
For $(\theta,\phi)=(\pi/3,\pi)$: $P(+x)\approx0.067$, $P(-x)\approx0.933$.

\medskip
\noindent\textbf{CC3 / PP3.} Full credit requires recognizing that increasing $N$ reduces the standard error by the expected $1/\sqrt{N}$ scaling and does not change the prepared state or the theoretical probability.

\medskip
\noindent\textbf{CC4 / PP4.} Full credit requires correct matrix products, correct identification of orthogonal resulting states, and the statement that orthogonal states cannot differ only by a global phase.

\medskip
\noindent\textbf{CC5 / PP5.} Full credit requires identifying the basis mismatch ($\ket{\psi}$ is expressed in the $x$ basis but the probability is claimed for a $z$-basis outcome) and either projecting directly with $|\braket{+z|\psi}|^2$ or first rewriting the state in the $z$ basis.

\end{document}